\documentclass[pre,amsmath,twocolumn,showpacs,superscriptaddress,nofootinbib]{revtex4-1}
\pdfoutput=1
\usepackage{graphicx}
\usepackage{amssymb}
\usepackage{amsmath}
\usepackage{epsfig}
\usepackage{color}
\usepackage{colortbl}

\makeatletter
   
   \def\CT@@do@color{%
      \global\let\CT@do@color\relax
      \@tempdima\wd\z@
      \advance\@tempdima\@tempdimb
      \advance\@tempdima\@tempdimc
   \advance\@tempdimb\tabcolsep
   \advance\@tempdima\tabcolsep
   \advance\@tempdima2\tabcolsep
            \kern-\@tempdimb
            \leaders\vrule
                    \hskip\@tempdima\@plus 1fill
            \kern-\@tempdimb
            \hskip-\wd\z@ \@plus -1fill }
\makeatother

\definecolor{Bblue}{rgb}{0.45,0.7,0.9}
\definecolor{Yyellow}{rgb}{0.9,0.8,0.3}

\newcommand{\dxt}{{\Delta}{\tilde x}}
\newcommand{\dvekx}{{\Delta}\vek{x}}
\newcommand{\dvekxt}{{\Delta}\vek{{\tilde x}}}
\newcommand{\dvekxht}{{\Delta}\vek{{\hat {\tilde x}}}}
\newcommand{\vek}[1]{\mbox{\boldmath$#1$}}
\newcommand{\mat}[1]{\mbox{\boldmath$#1$}}
\newcommand{\rem}[1]{}

\begin{document}

\title{Bayesian model selection with fractional Brownian motion}

\author{Jens Krog}
\affiliation{MEMPHYS, Department of Physics, Chemistry and Pharmacy,
University of Southern Denmark, DK-5230 Odense M, Denmark}
\author{Lars H. Jacobsen}
\affiliation{MEMPHYS, Department of Physics, Chemistry and Pharmacy,
University of Southern Denmark, DK-5230 Odense M, Denmark}
\author{Frederik W. Lund}
\affiliation{MEMPHYS, Department of Physics, Chemistry and Pharmacy,
University of Southern Denmark, DK-5230 Odense M, Denmark}
\author{Daniel W{\"u}stner}
\affiliation{Department of Biochemistry and Molecular Biology,
University of Southern Denmark, DK-5230 Odense M, Denmark}
\author{Michael A. Lomholt}
\affiliation{MEMPHYS, Department of Physics, Chemistry and Pharmacy,
University of Southern Denmark, DK-5230 Odense M, Denmark}

\begin{abstract}
We implement Bayesian model selection and parameter estimation for the case of fractional Brownian motion with measurement noise and a constant drift. The approach is tested on artificial trajectories and shown to make estimates that match well with the underlying true parameters, while for model selection the approach has a preference for simple models when the trajectories are finite. The approach is applied to observed trajectories of vesicles diffusing in Chinese hamster ovary cells. Here it is supplemented with a goodness-of-fit test, which is able to reveal statistical discrepancies between the observed trajectories and model predictions.
\end{abstract}


\date{\today}

\maketitle

\section{Introduction}

The nature of the motion of particles in biological cells is often found to deviate significantly from Brownian motion \cite{saxton97,golding06,weiss03,jeon11}. However, the most common method for analyzing the motion, estimation of the time-averaged mean square displacement (TA-MSD), cannot always distinguish whether the diffusion is non-Brownian. For example, the TA-MSD is linear in time for both Brownian motion and continuous time random walks with long tailed power law distributed waiting times \cite{lubelski08,he08}. Other estimators have been suggested, which are able to distinguish between some but not all classes of non-Brownian diffusion. Examples are $p$ variation tests \cite{magdziarz09}, first passage times \cite{condamin08}, and mean maximal excursions \cite{tejedor10}. See for instance \cite{meroz15} for a review. As further models of diffusion are introduced, the number and complextity of estimators increases, and it becomes unclear how to systematically compare the models once the data is limited. Is there a systematic approach?


In this article we approach the problem from a Bayesian perspective \cite{mackay03,jaynes03,sivia06}. Bayesian statistics provides a systematic framework for comparing different probabilistic models to select the one that best describes a given dataset. 
All subjectivity is clearly stated in the specification of the prior probability distributions for the model parameters. The framework relies on the computation of the \textit{likelihood} function, and can be applied on the raw trajectory, without filtering out information by computing other estimators. However, sometimes the computational complexity involved is an obstacle to a full Bayesian approach. We tackle this using the nested sampling approach introduced by Skilling \cite{skilling04}.

This work focuses on the specific model of anomalous diffusion called fractional Brownian motion (FBM) \cite{mandelbrot68}, in which the steps are correlated with long term memory leading to a MSD which is non-linear in time. In addition to using the Bayesian framework to distinguish pure and fractional Brownian motion, we also compare models featuring measurement noise and constant drift. Our implementation of the Bayesian inference is mainly implemented in Matlab while some calculations are implemented in C to increase performance. The scripts are publicly available at GitHub \cite{fbmgit}.

While the Bayesian approach provides rankings for a set of candidate models, it does not address the question of whether the best of these models actually describes the data well. We therefore supplement the Bayesian approach with a goodness-of-fit test in the form of the information content model check of \cite{krog17}, which calculates a $p$ value giving a measure of how typical the observed data is for a specific model.
 If extreme $p$ values are found, the analysis reveals that the corresponding model does not describe the data well, and that some important physical feature has been overlooked. There are other tests that one could have used alternatively: for instance a test of how the TA-MSD match the model prediction \cite{sikora17} or investigation of the detrending moving average statistic \cite{sikora18frac}. 
The information content check does not rely on the existence of an estimator like the MSD, but utilizes only the likelihood function, which is available from the Bayesian analysis.

A number of other works on single particle tracking from a Bayesian perspective already exist, e.g inference with hidden Markov models (HMM) \cite{das09,persson13,monnier15} and regular diffusion in a potential energy landscape \cite{masson09}. As all these approaches utilize the likelihood function and no other estimator, combining their analysis with ours is a straightforward exercise.
Such a combination, possible also with future works, would yield a systematic model selection among a large set of models for single particle tracking data. We also note that a number of previous works on the aspect of parameter estimation for FBM by Bayesian methods have been published previously, for instance \cite{beskos15,kozachenko15,meisam2017,meisam2017}.

The article is organized as follows. In Sec. \ref{Bayes_sec} we introduce Bayesian inference and the model of fractional Brownian motion with measurement errors and drift. The Nested sampling frameework enabling Bayesian inference for such models is also outlined.
We perform parameter estimation and model comparison for artificial and experimental data sets in Sec.~\ref{Perform_sec}, and then move on to supplement with independent goodness-of-fit tests in Sec.~\ref{sec:VIII}. We offer our conclusions on the results in Sec.~\ref{sec:IX}.
%
\section{Bayesian inference and fractional Brownian motion}\label{Bayes_sec}

When using Bayesian inference \cite{mackay03,jaynes03,sivia06} to select the most probable model among a number of models, $M_1$, $M_2$,\dots, based on some data the starting point is Bayes' formula
\begin{equation}\label{bayes}
P(M_i|\mathrm{data})=\frac{P(\mathrm{data}|M_i)P(M_i)}{P(\mathrm{data})}.
\end{equation}
The probability on the left hand side is the \textit{posterior} probability of the model given the data, while $P(\mathrm{data}|M_i)$ is the \textit{likelihood} or \textit{evidence} of the model and $P(M_i)$ is the \textit{prior} probability of the model. The model independent probability, $P(\mathrm{data})$, drops out when one compares the probability of different models by taking their ratio:
\begin{equation}\label{ratio}
\frac{P(M_{i}|\mathrm{data})}{P(M_{j}|\mathrm{data})}=
\frac{P(\mathrm{data}|M_{i})}{P(\mathrm{data}|M_{j})}\frac{P(M_{i})}{P(M_{j})}.
\end{equation}
If the models are considered equally probable before taking the data into account, i.e., $P(M_{i})=P(M_{j})$, the models are simply ranked by their evidence $Z_i=P(\mathrm{data}|M_i)$.

Usually a model includes a set of parameters $\vek{\theta}$, the values of which are to be inferred. For these parameters, a prior probability $\pi(\vek{\theta})=P(\vek{\theta}|M_i)$ must be assigned according to the knowledge (or lack of knowledge) of the parameters in the absence of the data. The evidence can then be calculated as an integral over the parameter space
\begin{equation}\label{evidence}
Z_i= \int \mathcal{L}(\vek{\theta})\pi(\vek{\theta})d\vek{\theta},
\end{equation}
where we have introduced the likelihood of the parameters $\mathcal{L}(\vek{\theta})= P(\mathrm{data}|\vek{\theta} ,M_i)$ as the probability of the data given the parameters of the considered model.

For a specific model $M_i$ the posterior probability distribution of the parameters:
\begin{equation}
\label{eq:posterior}
P(\vek{\theta}|M_i,\mathrm{data})=\frac{\mathcal{L}(\vek{\theta})\pi(\vek{\theta})}{Z_i}
\end{equation}
specifies the estimated parameter values and their uncertainties. In order to do so, however, we must be able to calculate the likelihood of the data, $\mathcal{L(\vek{\theta})}$. 

\subsection{The likelihood function for FBM}\label{sec:III}

The class of models analysed in this work is that of fractional Brownian motion. 
FBM in one dimension is a zero mean stationary Gaussian process where the displacements ${\tilde x}_k$ and ${\tilde x}_n$ from the starting point ${\tilde x}x_0 = 0$ at two later times $k\tau$ and $n\tau$  are correlated such that \cite{mandelbrot68}
\begin{equation}
\langle {\tilde x}_k{\tilde x}_n \rangle=D_H\left[(k\tau)^{2H}+(n\tau)^{2H}-|k\tau-n\tau|^{2H}\right].
\end{equation}
If the data consists of $N+1$ observations ${{\tilde x}_n : n=0,1,2,\dots,N}$ at evenly spaced time intervals of length $\tau$ then the corresponding one step displacements $\dxt_n={\tilde x}_n-{\tilde x}_{n-1}$ will have autocovariance function
\begin{align}
\gamma(k)&=\langle \dxt_{n}\dxt_{n+k}\rangle\nonumber\\
\label{eq:covk}
&= D_H\tau^{2H}\left[|k+1|^{2H}+|k-1|^{2H}-2|k|^{2H}\right].
\end{align}
Collecting the displacements in a column vector ${\dvekxt_N}$ with transpose ${\dvekxt}^T_N=[\dxt_1,\dxt_2,\ldots,\dxt_N]$ we can write the likelihood function, i.e., the probability of observing ${\dvekxt_N}$, as
\begin{equation}\label{like_FBM}
\mathcal{L}_x(\vek{\theta})=\frac{1}{(2\pi)^{N/2}|\mat{\Gamma}_N|^{1/2}}\exp\left(-\tfrac{1}{2}\dvekxt^T_N\mat{\Gamma}_N^{-1}\dvekxt_N\right)
\end{equation}
where $\mat{\Gamma}_N^{-1}$ is the inverse of the $N\times N$ covariance matrix with elements
\begin{equation}\label{gamma_inc}
\Gamma_{N,mn}=\langle \dxt_{m}\dxt_{n}\rangle=\gamma(m-n)
\end{equation}
and determinant $|\mat{\Gamma}_N|$. For a specific choice of $D_H$ and $H$, the likelihood function can then be calculated via Eq.~(\ref{gamma_inc})~and~(\ref{eq:covk}).

For a large data set, conventional inversion of $\mat{\Gamma}_N$ can be computationally demanding. We circumvent this difficulty by rewriting the likelihood expression by using the fact that
\begin{align}
P(\dvekxt_N|\vek{\theta})=&P(\dxt_N|\dvekxt_{N-1},\vek{\theta})\times P(\dvekxt_{N-1}|\vek{\theta})\nonumber\\
=&P(\dxt_N|\dvekxt_{N-1},\vek{\theta})
	\times P(\dxt_{N-1}|\dvekxt_{N-2},\vek{\theta})\nonumber\\
&\times\dots\times P(\dxt_1|\vek{\theta}),
\end{align}
where each conditional likelihood can be expressed as
\begin{equation}
P(\dxt_n|\dvekxt_{n-1},\vek{\theta})=\frac{1}{\sqrt{2\pi\sigma_n^2}}\exp\left(-\frac{(\dxt_n-\Delta{\tilde \mu}_n)^2}{2\sigma_n^2}\right).
\end{equation}
The mean $\Delta{\tilde \mu}_n$ and variance $\sigma_n^2$ in this expression are calculated iteratively using the Durbin-Levinson algorithm \cite{brockwell91}. Setting $\Delta{\tilde \mu}_1=0$ and $\sigma_1^2=\gamma(0)$ we iterate from $n = 1$ to $N-1$ by using
\begin{align}
\Delta{\tilde \mu}_{n+1}&=\sum_{j=1}^{n}\phi_{n,j}\dxt_{n+1-j},\label{eq:mnp1}\\
\sigma_{n+1}^2&=\sigma_{n}^2(1-\phi_{n,n}^2)\label{eq:snp1},
\end{align}
where
\begin{align}
\phi_{n,n}&= \frac{\gamma(n) - \sum_{j=1}^{n-1}\gamma(n-j)\phi_{n-1,j}}{\sigma_{n}^2},\label{eq:phinn}\\
\phi_{n,i}&=\phi_{n-1,i}-\phi_{n-1,n-i}\phi_{n,n} \,\mbox{  for  } 1\le i<n,\label{eq:phini}
\end{align}
with $\phi_{1,1}=\gamma(1)/\gamma(0)$. For completeness we include a derivation of Eqs. (\ref{eq:mnp1}-\ref{eq:phini}) in Appendix \ref{durbin-levinson}.

We have defined FBM for zero-mean processes, but we can straightforwardly relax this and allow for a constant average drift with velocity $v_x$. Labeling the actual measured displacements as $\Delta x_n$ we simply subtract the average trend by defining $\dxt_n=\Delta x_n-v_x\tau$.

In order to apply the model to single-particle tracking data obtained with a microscope producing two-dimensional images, we must generalize to two dimensions corresponding to two sets of coordinates; 
$({\tilde x}_n,{\tilde y}_n)$.
By assuming that the motion in the two dimension are independent, i.e. $\langle\tilde{x}_n \tilde{y}_n \rangle = 0$, the likelihood function for the two-dimensional fractional Brownian motions is given by
\begin{equation}
\mathcal{L}(D_H,H,v_x,v_y)=\mathcal{L}_x(D_H,H,v_x)\mathcal{L}_y(D_H,H,v_y),
\end{equation}
where $\mathcal{L}_y(D_H,H,v_y)=P(\Delta\vek{\tilde y}_N|D_H,H,v_y)$ is also given  
 by Eq. (\ref{like_FBM}).

\subsection{Measurement noise}\label{sec:IV}

When analysing single particle tracking data it can be necessary to distinguish between the experimentally observed particle position and the actual position, the difference being that the actual position can be obscured by inaccuracies in the measurement process \cite{savin05,berglund10}.

We include the possibility of a Gaussian measurement noise via the Durbin-Levinson algorithm. Labelling the actual particle $x$-positions as $x_n^{\rm clean}$ and the measurement noise for the positions $\eta_n$, the observed positions become
\begin{equation}
x_n=x_n^{\rm clean}+\eta_n.
\end{equation}

Assuming that the noise is memoryless with zero mean and variance $\langle \eta_n^2\rangle=\sigma_{\rm mn}^2$ we find that the displacements $\dxt_n=x_n-x_{n-1}-v_x\tau$ have autocovariance function $\gamma(k)=\langle \dxt_n\dxt_{n+k}\rangle$ with
\begin{equation}
\gamma(n)=\left\{\begin{array}{ll}\gamma^{\rm clean}(0)+2\sigma_{\rm mn}^2,  & \text{ for }n=0,\\
\gamma^{\rm clean}(1)-\sigma_{\rm mn}^2 & \text{ for } n=1,\\
\gamma^{\rm clean}(n)& \text{ for } n>1.\end{array}\right.
\end{equation}
where $\gamma^{\rm clean}(n)$ is the autocovariance in the absence of measurement noise corresponding to the underlying FBM. Thus the inclusion of this measurement noise is a simple modification of the Durbin-Levinson algorithm by the addition of one extra parameter, $\sigma_{\rm mn}$. 
It would be a straightforward matter to generalize the model of the noise to include correlations among the $\eta_n$ and calculate the corresponding autocovariance function $\gamma(n)$. Such correlations may arise due to the finite image acquisition time \cite{savin05,berglund10} or due to conformational changes of the observed objects .

\subsection{Prior distributions}\label{prior_distr}

The Bayesian formalism, outlined in Eqs.~(\ref{bayes})-(\ref{eq:posterior}), requires that a prior distribution $\pi(\vek{\theta})$ over the possible parameter values of the model is specified. The prior represents the knowledge, or lack of knowledge, about the system before the data is known. We will use uniform priors
\begin{equation}\label{uniform_prior}
\pi(\theta)=
\begin{cases} 
\frac{1}{\theta_{\rm max}-\theta_{\rm min}}, & \mbox{if   }\theta_{\rm min}\leq \theta \leq \theta_{\rm max} \\
0, & \mbox{otherwise}
\end{cases}
\end{equation}
for parameters such as $H$ that are unknown within some given interval from $\theta_{\rm min}$ to $\theta_{\rm max}$ and parameters, such as measurement noise or the components of the drift velocity, that may be zero. For the diffusion constant which is always positive, but where even the order of magnitude of the value could be unknown, we will use the socalled Jeffreys prior \cite{sivia06}
\begin{equation}\label{jeff_norm}
 \pi(\theta)=
\begin{cases} 
\frac{1}{\ln(\theta_{\rm max}/\theta_{\rm min})}\frac{1}{\theta},  & \mbox{for }\theta_{\rm min}\leq \theta \leq \theta_{\rm max} \\
0 & \mbox{otherwise}.
\end{cases}
\end{equation}
Since $D_H$ will change units when $H$ changes we will in practice use the deviation for a single step, which we will label $\sigma_H$, as the parameter for which we specify a Jeffrey's prior and then define $D_H=\sigma_H^2/(2\tau^{2H})$.

\subsection{Nested sampling}\label{NS_Sec}

Besides the calculation of the likelihood function, the main practical obstacle in Bayesian inference is the calculation of the evidence of a given model, since this often includes multidimensional integrals. We use the nested sampling procedure, as introduced by Skilling \cite{skilling04}, to accomplish this task. We briefly introduce the method and specific details of our implementation below, while we refer to \cite{sivia06} for further details. 
The essence of the nested sampling procedure is to rewrite Eq. (\ref{evidence}) on the form (suppressing the index $i$ of the model)
\begin{equation}\label{Z_simple}
Z=\int_0^1\, \mathcal{L}(X) dX.
\end{equation}
Here, the function $\mathcal{L}(X)$ (distinguished from $\mathcal{L}(\vek{\theta})$ by the variable name) is defined as the inverse of the function
\begin{equation}
X(\lambda)=\int_{\mathcal{L}(\vek{\theta})>\lambda}\pi(\vek{\theta})d\vek{\theta}.
\end{equation}
The variable $X$, which we will loosely call the prior mass, is the proportion of the parameter space with likelihood greater than $\lambda$. When implementing nested sampling, the integral in Eq. (\ref{Z_simple}) is estimated by the sum
\begin{equation}
Z\approx \sum_{i=1}^{i_{\rm max}}\mathcal{L}_i w_i\label{eq:Zapprox}.
\end{equation}
Here the likelihood values $\mathcal{L}_i$ are computed by generating $K$ `walkers' in the parameters space, denoted by $\vek{\theta}_k$, where $k=1,\dots,K$. The walkers are independently chosen points in parameter space, distributed randomly according to the prior $\pi(\vek{\theta})$ (In our implementaion we set $K=200$). 
From these $K$ walkers the one, $\vek{\theta}_k\equiv\vek{\theta}_{\rm min}^{(1)}$, with the smallest likelihood $\mathcal{L}(\vek{\theta}_k)\equiv\mathcal{L}_1$ is chosen for the $i=1$ term in Eq. (\ref{eq:Zapprox}). 

For $i>1$ the $\vek{\theta}_{\rm min}^{(i)}$ and $\mathcal{L}_i$ are generated iteratively in the same manner, except that the $K$ walkers are now restricted to be in the part of the parameter space where the likelihood is larger than $\mathcal{L}_{i-1}$. 
This means that for each new $i$ the considered prior mass will shrink on average by a factor $K/(K+1)$ since the $K$ walkers are distributed uniformly on the prior mass interval from $X=0$ to $X=X(\mathcal{L}_{i-1})$ with the largest value being $X(\mathcal{L}(\vek{\theta}_{\rm min}^{(i)}))$. Accordingly the weights $w_i$ in the sum are chosen to shrink as $w_i=w_{i-1}K/(K+1)$, with $w_1=1/(K+1)$ corresponding to the average prior mass discarded after the first iteration.

After each iteration we are left with $K-1$ walkers that satisfy the constraint of having likelihood larger than $\mathcal{L}_i=\mathcal{L}(\vek{\theta}_{\rm min}^{(i)})$. 
To supplement these with an independent sample, we choose one of the $K-1$ and duplicate it. 
The copy is then evolved through a random walk in parameter space to become independent of the other $K-1$ walkers.

Our procedure of evolving the walkers deviates slightly from the simplest one listed by Skilling, so we specify the technique below. 
Changing one parameter at a time, the walker is moved through parameter space as follows:
\begin{itemize}
\item{Defining $u=F(\theta)\equiv \int_{\theta_{\rm min}}^\theta \pi(\theta')d\theta'$, a $u^*$ is randomly chosen uniformly within the interval of length $l_u$ centered\footnote{We use periodic boundary conditions at $u^*=0$ and $u^*=1$, and initially $l_u=1$.} on $u$, and $\theta^*=F^{-1}(u^*)$ is calculated.}
\item{A trial position $\vek{\theta}^*$ in parameter space is constructed by supplementing $\theta^*$ with the current values of the other parameters for the walker. }
\item{If $\mathcal{L}(\vek{\theta}^*)>\mathcal{L}_i$ then the attempted move is accepted and the walker moves to this position in parameter space. If not, the move is rejected and the walker remains at the original position.}
\end{itemize}
The procedure is repeated for each parameter coordinate and the random walk then ends after $N_{\rm sweeps}=30$ attempted jumps for each parameter. To adjust the length of the moves as the prior mass shrinks we change the value of $l_u$ after each finished walk.
If $R$ denotes the fraction of rejected moves during the $N_{\rm sweeps}$ attempted, then we update
\begin{equation}
l_u \rightarrow \text{min}\left( l_u\,\text{exp}(0.25-R),\,1 \right)
\end{equation} 
 for that particular direction, which is then used for the next iteration.
 Our $l_u$ adjustment aims at obtaining a rejection fraction around $25\%$, while one could argue that the best fraction would be $50\%$. However, since the fraction of rejected steps is estimated from a walk possibly performed in a different area of parameter space than the following walk, we choose to be a bit conservative when updating $l_u$ such as to limit the number of walks where all moves in a specific direction are rejected.

Thus, the procedure differs from Skilling's in two aspects: (i) We consider each parameter on its own and store a separate $l_u$ for each, reflecting that some parameters are more tightly constrained by the likelihood requirement than others. (ii) The lengths are adjusted more conservatively, to decrease the risk of losing independence between the samples.

To obtain a termination criterion at some iteration $i=i_{\rm stop}$, we estimate the remainder of the integral as $Z_{\rm remain}=w_i\sum_{k=1}^K\mathcal{L}(\vek{\theta}_k)$. If $Z_{\rm remain}$ divided by the current estimate of the evidence $Z=\sum_{j=1}^i\mathcal{L}_j w_j$ is smaller than some fixed number (we have used $10^{-5}$) then we terminate the algorithm. At this point, we are left with $K$ walkers which are included in the evidence by adding the final $Z_{\rm remain}=w_{i_{\rm stop}}\sum_{k=1}^K \mathcal{L}(\vek{\theta}_k)$ to $Z$. In Eq. (\ref{eq:Zapprox}) we thus set $i_{\rm max}=i_{\rm stop}+K$, and define $\mathcal{L}_{i_{\rm stop}+k}=\mathcal{L}(\vek{\theta}_k)$ and $w_{i_{\rm stop}+k}=w_{i_{\rm stop}}$ for $k=1,\dots,K$.

With regards to the uncertainty on the estimated evidence we follow Skilling \cite{skilling04} and estimate the uncertainty of $\ln Z$ as $\sqrt{\mathcal{H}/K}$, where the information
\begin{equation}
\mathcal{H}=\int_0^1\frac{\mathcal{L}(X)}{Z}\ln\frac{\mathcal{L}(X)}{Z}d X
\end{equation}
is estimated as
\begin{equation}
\mathcal{H}\approx \sum_{i=1}^{i_{\rm max}} \frac{\mathcal{L}_i w_i}{Z}\ln\frac{\mathcal{L}_i}{Z}
\end{equation}

Finally the mean and variance of the parameters, or more generally the posterior average of any function $f(\vek{\theta})$, can be estimated as
\begin{equation}
\langle f(\vek{\theta})\rangle \approx \sum_{i=1}^{i_{\rm max}} f(\vek{\theta}^{(i)}_{\rm min})\frac{\mathcal{L}_i w_i}{Z}.\label{eq:postave}
\end{equation}

In case large data sets, we reduce the amount of stored samples $\vek{\theta}^{(i)}_{\rm min}$, to decrease the required computer memory. The number of samples have been kept below a threshold by letting a new sample, produced by the nested sampling algorithm, compete with a random old one. 
Of the pair, one is selected randomly with probabilities proportional to the samples's posterior probabilities. This sample is then assigned a new posterior probability equal to the sum of the two. This sum is then used instead of $\mathcal{L}_i w_i/Z$ in Eq. (\ref{eq:postave}) when calculating posterior averages.

%

\section{Parameter inference and model selection tests}\label{Perform_sec}

To demonstrate the strengths and weaknesses of the Bayesian approach, we have tested the implementation on a range of artificial trajectories as well as a bundle of experimental data where anomalous diffusion is suspected to appear.

In Table \ref{test_res_T1} the results of our analysis is shown for an artificially generated track. The underlying process is quite complex, and features positive correlations between steps ($H=3/4$) as well as some measurement noise, but zero drift.
\begin{table*}[t] 
\begin{center}
\begin{tabular}{ | r | r | r | r | r | r | r | r | r |} 
\hline
$i$ & $\log_{10}Z_i$ & $\sigma_H$ & $v_x\tau$ & $v_y\tau$ & $\sigma_{\rm mn}$ & $H$ & $\log_{10}\mathcal{L}_{\rm max}$ & $P(M_i|{\rm data})$
\\ \hline
1 & $-806.55 \pm 0.06$ & $24.9 \pm 0.9$ & $0$ & $0$ & $0$ & $1/2$ & $-804.64$ & $0.00000011$ \\
2 & $-804.20 \pm 0.12$ & $23.9 \pm 0.8$ & $9.9 \pm 1.7$ & $-2.5 \pm 1.8$ & $0$ & $1/2$ & $-797.01$ & $0.00002386$ \\
3 & $-809.14 \pm 0.09$ & $24.7 \pm 0.9$ & $0$ & $0$ & $1.8 \pm 1.3$ & $1/2$ & $-804.64$ & $0.00000000$ \\
4 & $-800.89 \pm 0.07$ & $24.6 \pm 0.9$ & $0$ & $0$ & $0$ & $0.626 \pm 0.025$ & $-797.98$ & $0.04930365$ \\
5 & $-806.67 \pm 0.14$ & $23.7 \pm 0.9$ & $9.9 \pm 1.7$ & $-2.5 \pm 1.7$ & $2.1 \pm 1.5$ & $1/2$ & $-797.01$ & $0.00000008$ \\
6 & $-803.63 \pm 0.13$ & $24.0 \pm 0.9$ & $9.6 \pm 2.7$ & $-2.7 \pm 2.7$ & $0$ & $0.587 \pm 0.029$ & $-795.46$ & $0.00008962$ \\
\rowcolor{Bblue}
7 & $-799.60 \pm 0.10$ & $20.0 \pm 1.9$ & $0$ & $0$ & $11.0 \pm 1.3$ & $0.799 \pm 0.058$ & $-794.50$ & $0.95046963$ \\
8 & $-803.53 \pm 0.13$ & $21.0 \pm 10.0$ & $8.8 \pm 14.2$ & $-2.7 \pm 13.9$ & $10.8 \pm 1.5$ & $0.793 \pm 0.079$ & $-793.87$ & $0.00011304$ \\
\hline
- & - & $20$ & $0$ & $0$ & $10$ & $0.75$ & $-794.70$ & - \\ \hline        
\end{tabular}
\end{center}
\caption{Results obtained by applying the nested sampling algorithm on a trajectory artificially generated with $N=200$ time steps and parameters as given by the last line in the table. The different models are labelled with $i$ and correspond to the 8 possible combinations of the following single point or uniform priors: $H=1/2$ or $0\le H \le 1$, $v_x=v_y=0$ or $-10^3\le v_x\tau,v_y\tau\le 10^3$, $\sigma_{\rm mn}=0$ or $0\le \sigma_{\rm mn}\le 10^3$. The prior on $\sigma_H$ is a Jeffreys prior with $\theta_{\rm min}=1$ and $\theta_{\rm max}=10^3$. The estimated parameter values for the broad priors are given with uncertainty (as mean $\pm$ standard deviation). $\mathcal{L}_{\rm max}$ is the maximal $\mathcal{L}_i$ found during the nested sampling run. The probability of the models are computed from the estimated mean of the evidences as $P(M_i|{\rm data})=Z_i/\sum_{j=1}^8 Z_j$, i.e., from Bayes' formula using $P(M_i)=1/8$ and $P({\rm data})=\sum_{j=1}^8 Z_j/8$.
}\label{test_res_T1}
\end{table*}     
As shown in the final column, the correct model ($i=7$) holds by far the largest evidence as expected. Between the 8 models, the true model is thus correctly identified in this case, while the simpler model ($i=4$), without measurement noise is the next best candidate.
The reason why the model without measurement noise does not yield an even smaller posterior probability lies within the broad prior on the measurement noise relative to the size of the true value. This broad prior reflects a large uncertainty in the model ($i=7$) with the measurement noise, which indeed weakens the model and thus decreases its evidence.
 Either increasing the length of the trajectory or narrowing the prior would lead to the true model being selected with even higher certainty (data not shown).

Table \ref{test_res_T2} shows the results for another simulated example, featuring negative correlations between the steps instead, as well as measurement noise and a constant drift in the $x$-direction. 
\begin{table*}[t]
\begin{center}
\begin{tabular}{ | c | c | c | c | c | c | c | c | c |} 
\hline
$i$ & $\log_{10}Z_i$ & $\sigma_H$ & $v_x\tau$ & $v_y\tau$ & $\sigma_{\rm mn}$ & $H$ & $\log_{10}\mathcal{L}_{\rm max}$ & $P(M_i|{\rm data})$
\\ \hline
1 & $-814.93 \pm 0.06$ & $26.1 \pm 0.9$ & $0$ & $0$ & $0$ & $1/2$ & $-813.01$ & $0.00000000$ \\
2 & $-814.43 \pm 0.12$ & $25.4 \pm 0.9$ & $0.2 \pm 1.7$ & $9.3 \pm 1.8$ & $0$ & $1/2$ & $-807.38$ & $0.00000000$ \\
3 & $-809.10 \pm 0.09$ & $17.7 \pm 1.2$ & $0$ & $0$ & $13.4 \pm 1.1$ & $1/2$ & $-804.78$ & $0.00000000$ \\
4 & $-815.47 \pm 0.08$ & $26.1 \pm 0.9$ & $0$ & $0$ & $0$ & $0.465 \pm 0.020$ & $-812.33$ & $0.00000000$ \\
5 & $-795.06 \pm 0.14$ & $8.9 \pm 1.1$ & $0.4 \pm 0.6$ & $9.3 \pm 0.6$ & $17.2 \pm 0.9$ & $1/2$ & $-784.71$ & $0.02679048$ \\
\rowcolor{Yyellow}
6 & $-793.53 \pm 0.15$ & $25.2 \pm 1.0$ & $0.3 \pm 0.2$ & $9.3 \pm 0.2$ & $0$ & $0.151 \pm 0.022$ & $-782.85$ & $0.90265055$ \\
7 & $-795.54 \pm 0.10$ & $7.4 \pm 2.2$ & $0$ & $0$ & $18.7 \pm 0.8$ & $0.919 \pm 0.044$ & $-790.59$ & $0.00892991$ \\
8 & $-794.70 \pm 0.15$ & $18.3 \pm 5.4$ & $0.4 \pm 0.4$ & $9.3 \pm 0.3$ & $10.9 \pm 5.1$ & $0.253 \pm 0.107$ & $-782.85$ & $0.06162905$ \\
\hline
- & - & $20$ & $0$ & $10$ & $10$ & $0.25$ & $-784.94$ & - \\ \hline        
\end{tabular}
\end{center}
\caption{Results obtained by applying the nested sampling algorithm on a subdiffusive trajectory generated with $N=200$ time steps and parameters as given by the last line in the table. The models and priors are the same as in Table \ref{test_res_T1}.
}\label{test_res_T2}
\end{table*}
In this case our analysis does not correctly distinguish between the negative correlations arising from the fractional Brownian nature and the similar contributions from the measurement noise. As before, a more narrow prior on the measurement noise or additional data can remedy this effect.
 It does not have problems discerning the drift though, although it offers as an unlikely alternative with around $1\%$ probability that the process is superdiffusive with larger measurement noise than the true value.

The lesson from the examples in Table~\ref{test_res_T1} and \ref{test_res_T2} is that the conclusions drawn with respect to model comparison are sensitive to the amount of data and the prior assignments. When the data becomes sparse or the prior widths increase, the analysis yields relatively higher evidende for the simpler models as compared to complex ones. It is thus crucial for model selection to assign priors corresponding to the uncertainty about each parameter taking \textit{everything except the data} into account.

In addition to the self consistency tests, the framework was utilized for a large ensemble of experimental data.
 Using a fluorescent analog of cholesterol enabled the tracking of sterol rich vesicles in chinese hamster ovarian cells, a system known to exhibit anomalous diffusion. For two different temperatures, an ensemble of vesicles were tracked, yielding 111 2-d trajectories at 25 $^{\circ}$C and 170 trajectories at 37 $^\circ$C \cite{lund12}.

\begin{figure}[h!]
\begin{center}
\includegraphics[width = 0.45\textwidth]{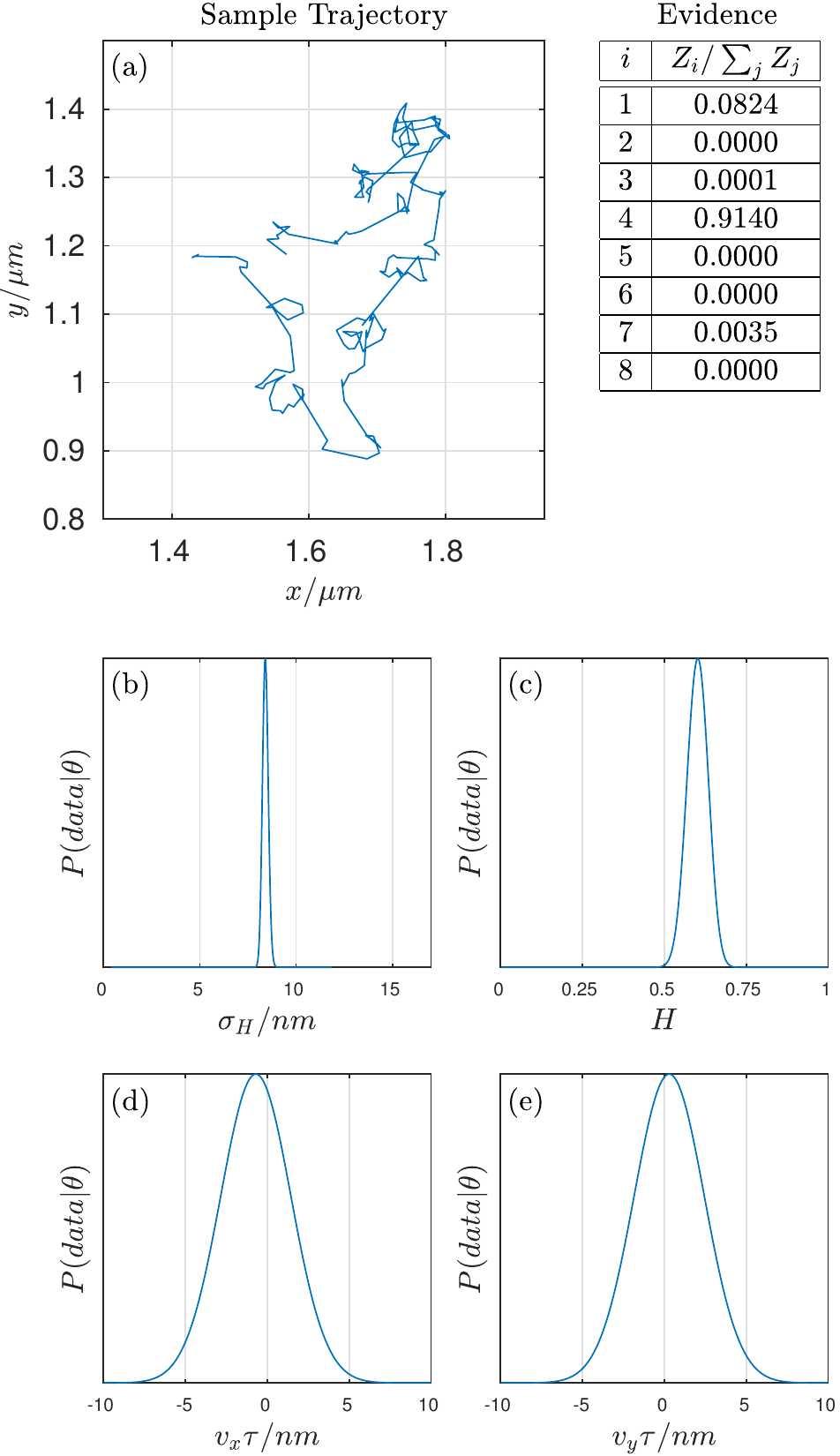}
\caption{
(a) A sample experimental trajectory besides the estimated evidences for each model. In (b)-(e) we display the evolution of the likelihood function around the maximum likelihood point for model $M_4$, i.e., pure fractional Brownian motion without drift.}\label{sample_like}
\end{center}
\end{figure}

We show a sample trajectory along with the evidence for each diffusion model and the behavior of the likelihood function around the maximum likelihood point for the best model in Fig.~\ref{sample_like}.
The model evidences are dominated by that of fractional Brownian motion, and indeed the Hurst parameter is estimated to be $H=0.605\pm0.035$, quite incompatible with pure Brownian motion at $H_{\text{pure}}=1/2$.
Note that a key feature of the Bayesian framework is that it yields probability distributions about each inferred parameter, and not just a mean and error, although these are readily available. For completeness, we demonstrate the complete parameter mean and error outputs of the analysis in Table~\ref{test_res_T3}.
\begin{table*}
\begin{center}
    \begin{tabular}{ | l | l | l | l | l | l | l | l | l |} 
    \hline
$i$ & $\log_{10}Z_i$ & $\sigma_H$ & $v_x\tau$ & $v_y\tau$ & $\sigma_{\rm mn}$ & $H$ & $\log_{10}\mathcal{L}_{\rm max}$ & $P(M_i|{\rm data})$\\ \hline
1 & $-742.78 \pm 0.06$ & $17.6 \pm 0.6$ & $0$ & $0$ & $0$ & $1/2$ & $-740.84$ & $0.04846667$ \\
2 & $-748.34 \pm 0.12$ & $17.6 \pm 0.6$ & $-0.5 \pm 1.2$ & $-0.0 \pm 1.2$ & $0$ & $1/2$ & $-740.79$ & $0.00000013$ \\
3 & $-745.39 \pm 0.09$ & $17.6 \pm 0.6$ & $0$ & $0$ & $1.1 \pm 0.8$ & $1/2$ & $-740.84$ & $0.00011981$ \\
4 & $-741.49 \pm 0.07$ & $17.7 \pm 0.6$ & $0$ & $0$ & $0$ & $0.605 \pm 0.035$ & $-738.69$ & $0.94665709$ \\
5 & $-750.93 \pm 0.14$ & $17.5 \pm 0.7$ & $-0.6 \pm 1.3$ & $0.0 \pm 1.3$ & $1.1 \pm 0.9$ & $1/2$ & $-740.79$ & $0.00000000$ \\
6 & $-747.22 \pm 0.13$ & $17.9 \pm 0.7$ & $-0.7 \pm 2.5$ & $0.5 \pm 2.6$ & $0$ & $0.615 \pm 0.036$ & $-738.66$ & $0.00000177$ \\
7 & $-743.79 \pm 0.10$ & $17.5 \pm 0.7$ & $0$ & $0$ & $2.0 \pm 1.4$ & $0.618 \pm 0.039$ & $-738.69$ & $0.00475450$ \\
8 & $-748.93 \pm 0.14$ & $17.6 \pm 0.8$ & $-0.8 \pm 2.9$ & $0.4 \pm 2.8$ & $2.3 \pm 1.5$ & $0.641 \pm 0.044$ & $-738.66$ & $0.00000003$ \\
\hline
\end{tabular}
\end{center}
\caption{Results obtained by applying the nested sampling algorithm on the trajectory of a vesicle in a cell measured by fluorescence microscopy. The positions are measured in nanometers with a time step $\tau=0.5$ second. The models and priors are the same as in Table \ref{test_res_T3}.
}\label{test_res_T3}
\end{table*}

\begin{figure}
\begin{center}
\includegraphics[width = 0.45\textwidth]{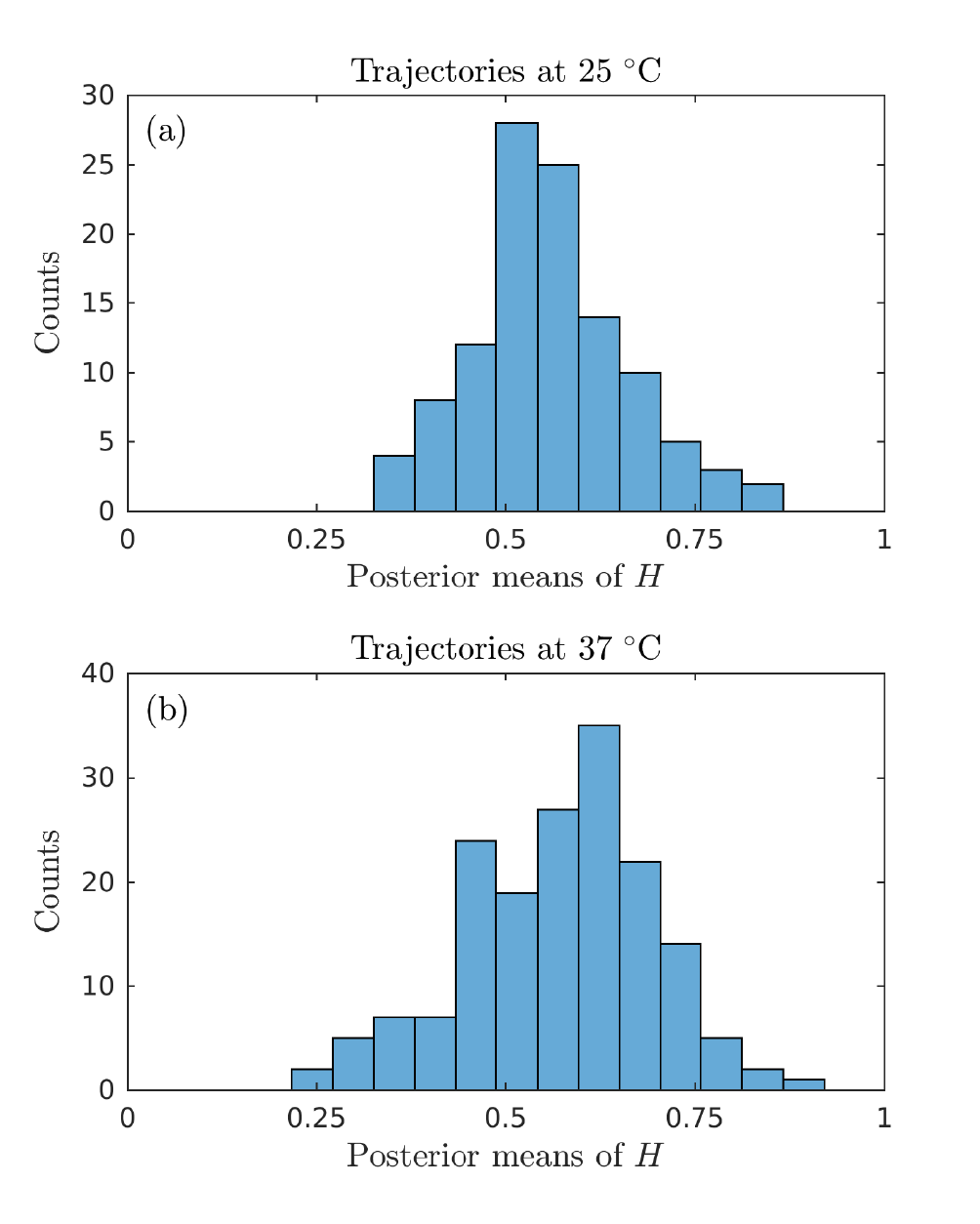}
\caption{Histograms of the posterior mean of $H$ for each experimental trajectory. In (a) results for trajectories captured at 25 $^{\circ}$C are shown, while (b) depicts the distribution for trajectories captured at 37 $^{\circ}$C}
\label{hurst_hist}
\end{center}
\end{figure}

To evaluate the degree of superdiffusion in the ensemble, we calculate the mean of the posterior distribution for each trajectory, and combine them the histograms in Fig.~\ref{hurst_hist}. A comparison between the histograms, indicate that the superdiffusion is more profound at 37 $^{\circ}$C, as also found in \cite{lund12}.

\begin{figure}
\begin{center}
\includegraphics[width = 0.45\textwidth]{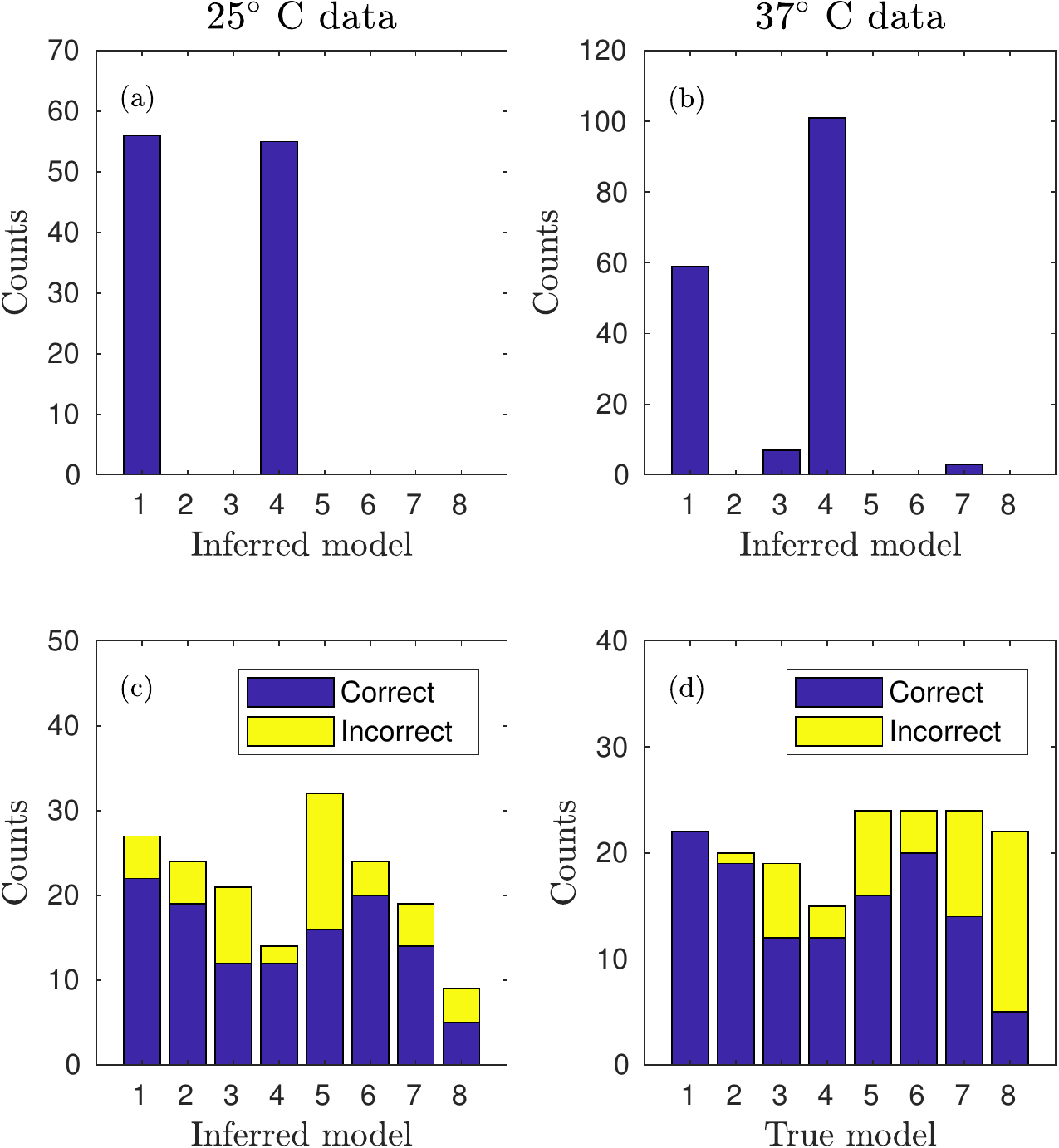}
\caption{
(a): Histogram of the most probable model for each trajectory for the data at $25^\circ$. (b): corresponding histogram for data at $37^\circ$. (c): corresponding histogram for 170 artificial trajectories generated using parameters drawn from to the priors given in Table \ref{test_res_T1} with a $1/8$ probability for each model. The top part of the bars indicates the wrongly selected models. (d): histogram of the corresponding true models with the top part indicating how many were wrongly selected by the Bayesian inference approach.
}\label{fig2}
\end{center}
\end{figure}

Turning to the model selection aspect of the Bayesian framework, we compare the model selection results from the experimental data with those of an ensemble of artificial trajectories. To mirror the experimental data, 170 artifical trajectories were simulated, where the underlying model was chosen with equal probability from the 8 possible models. Model parameters were then drawn from the prior distributions given below Table~\ref{test_res_T1}. The model selection procedure was then performed as for the experimental data with nested sampling, and the results are displayed in Fig.~\ref{fig2}.

As shown in the upper left of Fig. \ref{fig2} then the nested sampling algorithm selected Model 1 (pure Brownian motion) as the most probable for about half of the trajectories at 25 $^\circ$C, while Model 4 (fractional Brownian motion) was selected for the other half. 
The distribution in the upper right of Fig.~\ref{fig2} shows that selection of fractional Brownian motion is more pronounced at 37$^\circ$C, confirming the previous results . 
In the lower left plot we illustrate the results of a test measuring the model selection effectiveness on the artificial data. The bars indicate how many times each model was selected as the most probable, the lower part indicating correct choices, while the upper part represents incorrect ranking. The correct model was ranked highest 123 out of 170 times. 
The reason for the discrepancy is indicated in the histogram in the lower right, where the x-axis represents the true underlying model, while the bars indicate how often it was ranked highest. The plot indicates the well-known fact, that the Bayesian approach includes an Occam's razor effect: data from the simple model 1 was correctly classified each time, while the data from the complex model 8 was quite often appointed to one of the simpler models. This can also be viewed as an instance of the Jeffreys-Lindley paradox \cite{cousins17}.
\begin{figure}
\begin{center}
\includegraphics[width = 0.45\textwidth]{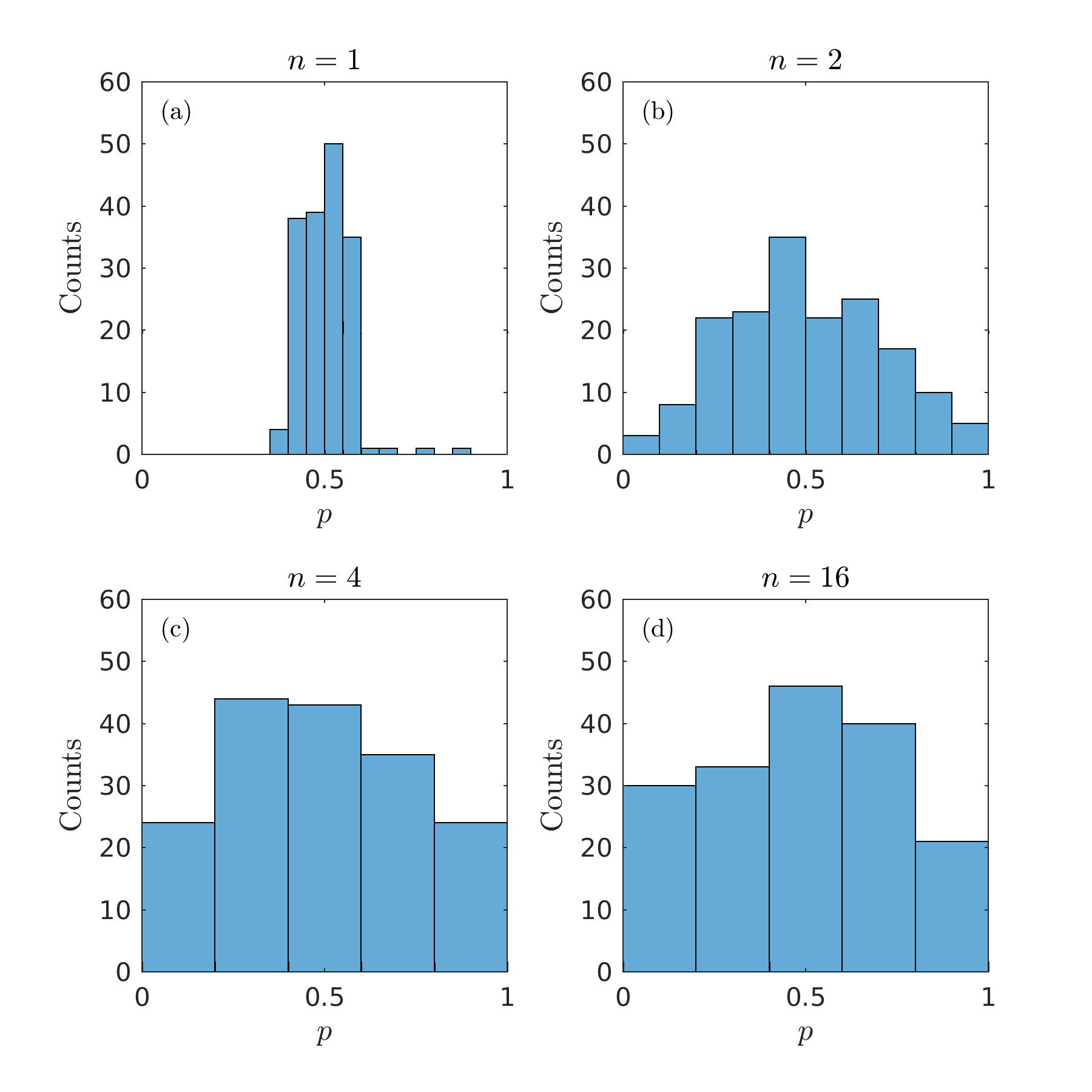}
\caption{Histograms of the $p$-values defined in Eq. (\ref{eq:pn}) for the same artificial trajectories as used in the lower plots of Fig. \ref{fig2}. The values are for $n=1$ (top left), $n=2$ (top right), $n=4$ (bottom left) and $n=16$ (bottom right) and the model with the highest evidence for the trajectory.}\label{fig3}
\end{center}
\end{figure}
\begin{figure}
\begin{center}
\includegraphics[width = 0.45\textwidth]{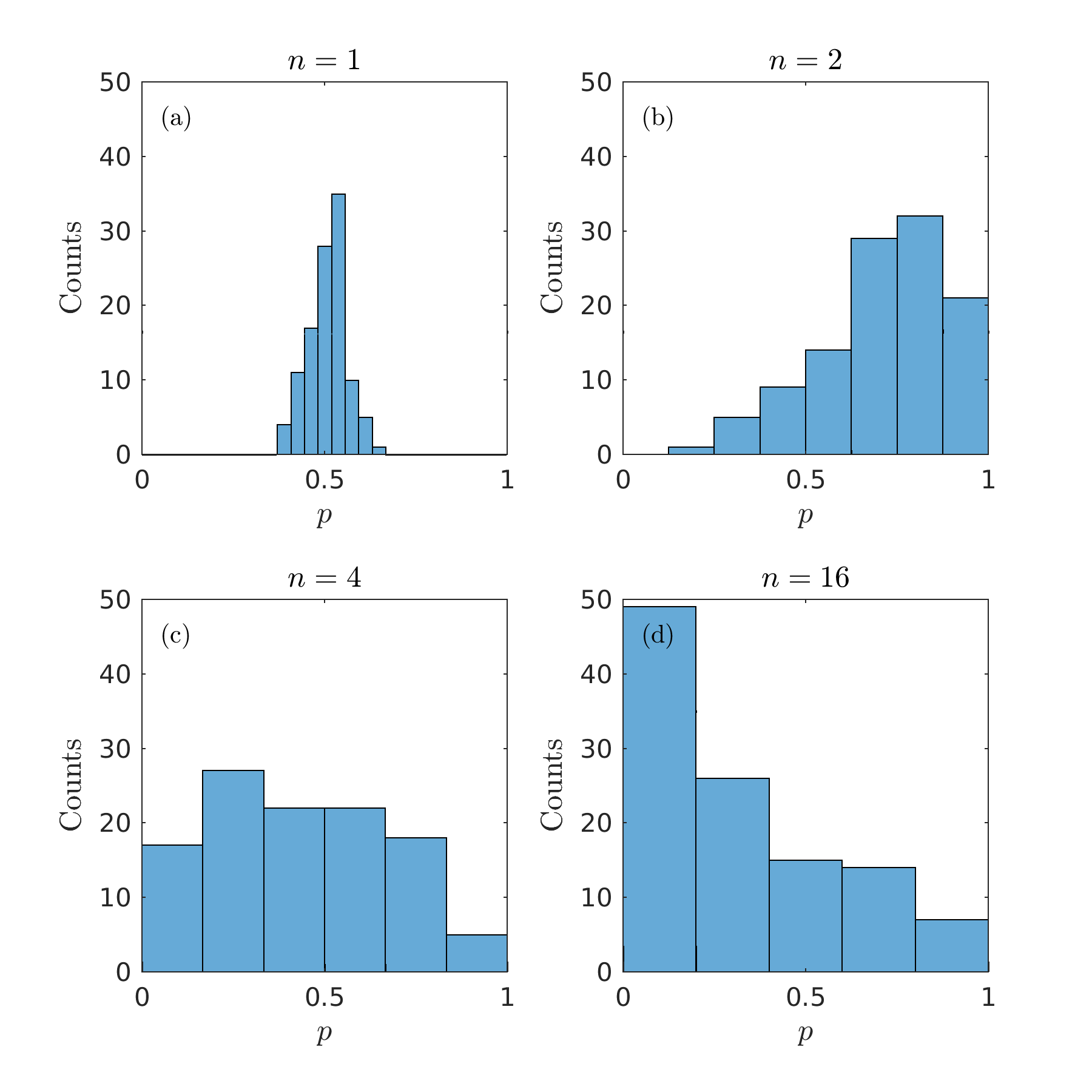}
\caption{Same as Fig. \ref{fig3} but for the experimental data at $25^\circ$.}\label{fig4}
\end{center}
\end{figure}

\section{Goodness-of-fit test}\label{sec:VIII}

The main shortcoming of the Bayesian analysis is, that it yields only relative model rankings and parameter values for the candidate models. To investigate how well the models describe the data, we supplement the Bayesian analysis with a goodness-of-fit test. This test checks whether the probability of the observed trajectory $\dvekx_N$ is typical for the inferred model. 
 We take our starting point in the $p$ value \cite{krog17}
\begin{equation}
p=\langle \Theta[P(\dvekx^*_N |\vek{\theta}^*,M_i)-P(\dvekx_N |\vek{\theta}^*,M_i)] \rangle,
\end{equation}
where $\Theta[\cdot]$ is the Heaviside step function and the average is over the joint posterior distribution $P(\dvekx^*_N,\vek{\theta}^*|M_i)$ for the parameters $\vek{\theta}^*$ and hypothetical trajectories $\dvekx^*_N$.

For each model $M_i$, we estimate $p$ by drawing $N_{\rm repl}=100$ sets of parameter values $\vek{\theta}^*_k$, $k=1,\cdots,N_{\rm repl}$ among the samples $\vek{\theta}^{(i)}_{\rm min}$ according to their posterior probabilities $\mathcal{L}_i w_i/Z$. For each of the chosen parameter samples a replicated trajectory $\dvekx^*_N$ is generated according to the corresponding stochastic model and then the $p$ value is estimated by averaging $\Theta[P(\dvekx^*_N |\vek{\theta}^*,M_i)-P(\dvekx_N |\vek{\theta}^*,M_i)]$ over the $N_{\rm repl}$ replicated trajectories.

The reasoning of such a $p$-value is as follows: If the model $M_i$ and parameters $\vek{\theta}^*$ are a perfect description of the data, then the probability distribution for the $p$-value would be uniform between zero and one. Thus, if the $p$-value turns out to be improbably close to the extremities zero or one, then one can conclude that the model probably does not describe the data well. However, as proven in an appendix of \cite{krog17}, the $p$ value as calculated above for the full trajectory will be valued close to $0.5$ for the type of diffusive models studied here no matter how badly it otherwise describes the data. The reason for this is that the data which is used for the goodness-of-fit is the same data that has already been used for the parameter inference. The test is therefore extended by considering how well the model describes the data when considering longer steps in time. Thus, the estimator used for the check is different than that which served as score for the parameter inference. The longer time steps are achieved simply by removing data points, keeping only every $n$th position along the trajectory for some integer $n$. We thus construct the steps in the $x$-direction of a trajectory $\Delta\vek{x}^{(n)}_{N'}$ by keeping only the $x$-coordinates $x^{(n)}_i=x_{n i}$ for the possible integers $i=0,\dots,N'$ and construct the corresponding steps $\Delta x^{(n)}_{i}=x^{(n)}_{i}-x^{(n)}_{i-1}$. The replicated trajectories $\dvekx^{(n)*}_N$ are constructed similarly, and the $p$-values are estimates of
\begin{equation}
p=\langle \Theta[P(\dvekx^{(n)*}_N |\vek{\theta}^*,M_i)-P(\dvekx^{(n)}_N |\vek{\theta}^*,M_i)] \rangle\label{eq:pn},
\end{equation}
obtained in the same way as described above but using the time step $n\times\tau$.

For the artificial trajectories corresponding to the lower plots in Fig. \ref{fig2} the distribution of the $p$-values is shown in Fig.~\ref{fig3}. Note how the values cluster around $p=0.5$ for $n=1$, but becomes increasingly uniform as $n$ is increased. As a reference, the experimental data at $25^\circ$ Celsius give rise to a different pattern, see Fig. \ref{fig4}. When $n$ become larger than one, the $p$-values do not spread out uniformly. At $n=2$ the distribution moves towards $p=1$, which means that trajectories replicated from the models tend to be more probable than the observed trajectories. As $n$ is increased even further, the $p$-values distribution tend towards $p=0$ instead, which indicates that the experimental data is more probable than what is expected for the model. An explanation of this behavior could be, that the correlations between steps at short and long time scales is different from that described by a simple power law. Thus a reasonable conclusion from the $p$-values is that the set of candidate models should be expanded to achieve models that describe the data better. Note that this conclusion could not have been reached by just considering the evidences of the candidate models, since they only tell you how to rank the candidate models relative to each other. The corresponding plots at $37^\circ$ are qualitatively similar to Fig. \ref{fig4}.

\section{Conclusion}\label{sec:IX}

We implemented Bayesian model selection and parameter estimation for fractional Brownian motion with  drift and measurement noise. The approach was tested on artificial trajectories and found to make estimates of the parameters and uncertainties that are consistent with the true underlying parameters. With limited data the Bayesian approach was able to select the true underlying model, except when the deviation from a simpler model is small compared with the broadness of the prior.

For the analysis of experimental data from vesicles tracked in Chinese hamster ovarian cells, the approach favored both regular and anomalous diffusion, with an increased tendency towards superdiffusion for intact cells, consistent with previous findings\cite{lund12}. However, a model check regarding the typicality of the steps observed at different time scales, revealed shortcomings of the fractional Brownian motion model.
We observed tendencies for the experimental trajectories to be untypically improbable at short time-scales and untypically probable at longer time-scales. This suggests that the set of candidate models does not fully cover the kind of stochastic process that governs the data, and thus that the set of candidate models should be expanded for these particular data sets. We note that the present approach could easily be modified to include models of stationary Gaussian processes with different autocovariance functions than the one for FBM. Thus we expect that the approach can be applied to single particle tracking in many different situations.

\section{Acknowledgement}

JK and MAL acknowledge funding from Danish Council for Independent Research - Natural Sciences (FNU), grant number 4002-00428B.

\appendix

\section{Durbin-Levinson algorithm}\label{durbin-levinson}

Our starting for the derivation of the Durbin-Levinson algorithm is an expression of $\dxt_{n+1}$ given all the previous steps
\begin{align}
\dxt_{n+1}&=\sum_{j=1}^{n}\phi_{n,j}\dxt_{n+1-j}+z_{n+1}\nonumber\\
&=\vek{\Phi}_{n,n}^T\dvekxht_n+z_{n+1}\label{fGn_process_short}
\end{align}
Here the sequence of coefficients $\vek{\Phi}_{n,n}^T=[\phi_{n,1},\phi_{n,2},\ldots,\phi_{n,n}]$ controls the correlation with the previous steps, the hat means that the order of the elements in the vector has been reversed: $\dvekxht_n^T=[\dxt_{n},\dxt_{n-1},\ldots,\dxt_{1}]$, and $z_{n+1}$ is a zero mean Gaussian noise which is independent of the previous displacements $\dvekxt_n$.

The coefficients $\phi_{n,i}$ can be determined by multiplying both sides of Eq. (\ref{fGn_process_short}) by $\dxt_i$, $i<n+1$ and averaging to get
\begin{equation}
\gamma(n+1-i)=\sum_{j=1}^n\phi_{n,j}\gamma(n+1-j-i)
\end{equation}
With $\vek{c}_{n}^T=[\gamma(1), \gamma(2), \ldots , \gamma(n)]$ this equation can be written in matrixform as $\boldsymbol{\Gamma}_n\vek{\Phi}_{n,n}=\mathbf{c}_n$ or
\begin{equation}\label{YW_inv}
\boldsymbol{\Phi}_{n,n}=\boldsymbol{\Gamma}_n^{-1}\mathbf{c}_n 
\end{equation}
In principle this determines the coefficients $\phi_{n,i}$ in terms of the autocovariance function $\gamma(n)$. However, it involves the inverse of $\Gamma_N$, which we would like to avoid calculating. Instead we aim at obtaining an iterative procedure for calculating $\phi_{n,i}$ step by step. 

Before going further with that let us find expressions for the conditional mean
\begin{equation}
\Delta{\tilde \mu}_{n+1}=\int \dxt_{n+1} P(\dxt_{n+1}|\dxt_1,\dots,\dxt_n)d\dxt_{n+1}
\end{equation}
of $\dxt_{n+1}$ given $\dxt_1,\dots,\dxt_n$
and corresponding variance $\sigma^2_{n+1}$. For the conditional mean we see directly from Eq. (\ref{fGn_process_short}) that
\begin{equation}\label{mu_derive}
\Delta{\tilde \mu}_{n+1}=\vek{\Phi}_{n,n}^T\dvekxht_n
\end{equation}
which is Eq. (\ref{eq:mnp1}).
From Eq. (\ref{fGn_process_short}) we also see that
the variance $\sigma_{n+1}^2$ of $\dxt_{n+1}$ given $\dxt_1,\dots,\dxt_n$ is the same as the variance of $z_{n+1}$. Thus we find
\begin{align}
\sigma_{n+1}^2 &=\langle z_{n+1}^2 \rangle=\langle (\dxt_{n+1}-\vek{\Phi}_{n,n}^T\dvekxht_n )^2\rangle\\
&= \gamma(0)-2 \vek{\Phi}_{n,n}^T\vek{c}_n+\vek{\Phi}_{n,n}^T \langle \dvekxht_n\dvekxht_n^T \rangle \vek{\Phi}_{n,n}\\
&=\gamma(0)- \vek{\Phi}_{n,n}^T\vek{c}_n\label{conditional_var}
\end{align}

Let us now return to finding iterative expressions for $\phi_{n,i}$. First note that 
$\boldsymbol{\Gamma}_{n+1}$ can be written in a block structure
\begin{equation}
\mat{\Gamma}_{n+1} = 
\begin{bmatrix}
       \mat{\Gamma}_{n} & \hat{\vek{c}}_{n}   \\[0.3em]
       \hat{\vek{c}}_{n}^T & \gamma(0) \\[0.3em]
     \end{bmatrix}
\end{equation}
where $\hat{\vek{c}}_{n}^T=[\gamma(n), \gamma(n-1), \ldots , \gamma(1)]$. This we can use to write
$\boldsymbol{\Gamma}_{n+1}\boldsymbol{\Phi}_{n+1,n+1}=\mathbf{c}_{n+1}$
as
\begin{equation}
\begin{bmatrix}
       \boldsymbol{\Gamma}_n & \hat{\boldsymbol{c}}_n   \\[0.3em]
       \hat{\boldsymbol{c}}_{n}^T & \gamma(0) \\[0.3em]
     \end{bmatrix}\left[ \begin{array}{c} \boldsymbol{\Phi}_{n+1,n}\\ \phi_{n+1,n+1}\end{array} \right]=\left[ \begin{array}{c} \boldsymbol{\mathbf{c}}_n\\\gamma(n+1)\end{array} \right]
\end{equation}
where $\vek{\Phi}_{n+1,n}^T=(\phi_{n+1,1},\phi_{n+1,2},\ldots,\phi_{n+1,n})$.
Writing this out as two equations we obtain
\begin{subequations}
\begin{align}
\boldsymbol{\Gamma}_n\boldsymbol{\Phi}_{n+1,n}+\hat{\boldsymbol{c}}_n\phi_{n+1,n+1} &= \mathbf{c}_n\label{eqn: a 1}\\
\hat{\boldsymbol{c}}_{n}^T\boldsymbol{\Phi}_{n+1,n}+\gamma(0)\phi_{n+1,n+1} &= \gamma(n+1) \label{eqn: b 2} 
\end{align}
\end{subequations}
This can be rearranged to find
\begin{subequations}
\begin{align}
\boldsymbol{\Phi}_{n+1,n}=\boldsymbol{\Gamma}_n^{-1}(\mathbf{c}_n-\hat{\boldsymbol{c}}_n\phi_{n+1,n+1})&\label{eqn: aa 1}\\
\hat{\boldsymbol{c}}_{n}^T\mat{\Gamma}_n^{-1}(\mathbf{c}_n-\hat{\boldsymbol{c}}_n\phi_{n+1,n+1})+\gamma(0)\phi_{n+1,n+1} &= \gamma(n+1) \label{eqn:bb 2} 
\end{align}
\end{subequations}
and further exploiting Eq. (\ref{YW_inv}) to obtain
\begin{subequations}
\begin{align}
\boldsymbol{\Phi}_{n+1,n}&=\boldsymbol{\Phi}_{n,n}-\hat{\boldsymbol{\Phi}}_{n,n}\phi_{n+1,n+1}\label{eqn: aaa 1}\\
\phi_{n+1,n+1}&=\frac{\gamma(n+1)-\hat{\boldsymbol{c}}_n^T\boldsymbol{\Phi}_{n,n}}{\gamma(0)-\hat{\boldsymbol{c}}_n^T\mat{\Gamma}_n^{-1}\hat{\mathbf{c}}_n}\label{eqn: bbb 2}
\end{align}
\end{subequations}
Since $\hat{\boldsymbol{c}}_n^T\Gamma_n^{-1}\hat{\mathbf{c}}_n=\boldsymbol{c}_n^T\Gamma_n^{-1}\mathbf{c}_n$ we see that the denominator of Eq. (\ref{eqn: bbb 2}) is the variance given in Eq. (\ref{conditional_var}). Thus we arrive at (after shifting the index $n$ by 1)
\begin{subequations}
\begin{align}
\phi_{n,i}&=\phi_{n-1,i}-\phi_{n-1,n-i}\phi_{n,n} \mbox{ for } i<n\\
\phi_{n,n}&= \frac{\gamma(n) - \sum_{j=1}^{n-1}\gamma(n-j)\phi_{n-1,j}}{\sigma_{n}^2}
\end{align}
\end{subequations}
which gives the iterative procedure described in Eqs. (\ref{eq:phinn}) and (\ref{eq:phini}).

Returning to the variance we can now simplify it using Eqs. (\ref{eqn: aaa 1}) and (\ref{eqn: bbb 2}) to   
\begin{eqnarray}
\sigma_{n+1}^2&=&\gamma(0)-\boldsymbol{c}_{n}^T\boldsymbol{\Phi}_{n,n}\nonumber\\ 
&=&\gamma(0)-[\mathbf{c}_{n-1}^T,\gamma(n)]\begin{bmatrix}\boldsymbol{\Phi}_{n,n-1}\\ \phi_{n,n}
\end{bmatrix}\nonumber\\
&=&\gamma(0)-(\mathbf{c}_{n-1}^T\boldsymbol{\Phi}_{n,n-1}+\gamma(n)\phi_{n,n})\nonumber\\
&=&\gamma(0)-(\mathbf{c}_{n-1}^T(\boldsymbol{\Phi}_{n-1,n-1}-\hat{\boldsymbol{\Phi}}_{n-1,n-1}\phi_{n,n})\nonumber\\
& &+\gamma(n)\phi_{n,n})\nonumber\\
&=&\sigma_{n}^2-\sigma_{n}^2\phi_{n,n}\frac{\gamma(n)-\mathbf{c}_{n-1}^T\hat{\boldsymbol{\Phi}}_{n-1,n-1}}{\sigma_{n}^2}\nonumber\\
&=& \sigma_{n}^2(1-\phi_{n,n}^2)\label{sigma_derive}
\end{eqnarray}
which is Eq. (\ref{eq:snp1}). This concludes our derivation of the Durbin-Levinson algorithm.

\bibliography{fbm_refs}

\end{document}